# Biometric security technology[1]

Marcos Faundez-Zanuy
Escola Universitaria Politècnica de Mataró
Avda. Puig i Cadafalch 101-111
08303 MATARO (BARCELONA) SPAIN
E-mail: faundez@eupmt.es http:www.eupmt.es/veu

**ABSTRACT**
This paper presents an overview of the main topics related to biometric security technology, with the main purpose to provide a primer on this subject.
Biometrics can offer greater security and convenience than traditional methods for people recognition. Even if we do not want to replace a classic method (password or handheld token) by a biometric one, for sure, we are potential users of these systems, which will even be mandatory for new passport models. For this reason, to be familiarized with the possibilities of biometric security technology is useful.

## INTRODUCTION

The word Biometrics comes from the Greek words "bios" (life) and "metrikos" (measure). Strictly speaking, it refers to a science involving the statistical analysis of biological characteristics. Thus, we should refer to biometric recognition of people, as those security applications that analyze human characteristics for identity verification or identification. However, we will use the short term "biometrics" to refer to "biometric recognition of people".
Biometric recognition offers a promising approach for security applications, with some advantages over the classical methods, which depend on something you have (key, card, etc.), or something you know (password, PIN, etc.). A nice property of biometric traits is that it is based on something you are or something you do, so you do not need to remember anything neither to hold any token.
Authentication methods by means of biometrics are a particular portion of security systems, with a good number of advantages over classical methods. However, there are also drawbacks (see table 1).

| Authentication method | Advantages | Drawbacks |
|---|---|---|
| Handheld tokens (card, ID, passport, etc.) | ▪ A new one can be issued.<br>▪ It is quite standard, although moving to a different country, facility, etc. | ▪ It can be stolen.<br>▪ A fake one can be issued.<br>▪ It can be shared.<br>▪ One person can be registered with different identities. |
| Knowledge based (password, PIN, etc.) | ▪ It is a simple and economical method.<br>▪ If there are problems, it can be replaced by a new one quite easily. | ▪ It can be guessed or cracked.<br>▪ Good passwords are difficult to remember.<br>▪ It can be shared.<br>▪ One person can be registered with different identities. |
| Biometrics | ▪ It cannot be lost, forgotten, guessed, stolen, shared, etc.<br>▪ It is quite easy to check if one person has several identities.<br>▪ It can provide a greater degree of security than the other ones. | ▪ In some cases a fake one can be issued.<br>▪ It is neither replaceable nor secret.<br>▪ If a person's biometric data is stolen, it is not possible to replace it. |

**Table 1. Advantages and drawbacks of the three main authentication method approaches.**

Depending on the application, one of the previous methods, or a combination of them, will be the most appropriate. This paper describes the main issues to be known for decision making, when trying to adopt a biometric security technology solution.

## BIOMETRIC TRAITS

The first question is: Which characteristic can be used for biometric recognition? As common sense says, a good biometric trait must accomplish a set of properties. Mainly they are [1]:

- Universality: Every person should have the characteristic.

[1] This work has been supported by FEDER and MEC, TIC-2003-08382-C05-02

- Distinctiveness: Any two persons should be different enough to distinguish each other based on this characteristic.
- Permanence: the characteristic should be stable enough (with respect to the matching criterion) along time, different environment conditions, etc.
- Collectability: the characteristic should be acquirable and quantitatively measurable.
- Acceptability: people should be willing to accept the biometric system, and do not feel that it is annoying, invasive, etc.
- Performance: the identification accuracy and required time for a successful recognition must be reasonably good.
- Circumvention: the skill of fraudulent people and techniques to fool the biometric system should be negligible.

Biometric traits can be split into two main categories:

**Physiological biometrics:** it is based on direct measurements of a part of the human body. Fingerprint, face, iris and hand-scan recognition belong to this group.

**Behavioral biometrics:** it is based on measurements and data derived from an action performed by the user, and thus indirectly measures some characteristics of the human body. Signature, gait, gesture and key stroking recognition belong to this group.

However, this classification is quite artificial. For instance, the speech signal depends on behavioral traits such as semantics, diction, pronunciation, idiosyncrasy, etc. (related to socio-economic status, education, place of birth, etc.) [2]. However, it also depends on the speaker's physiology, such as the shape of the vocal tract. On the other hand, physiological traits are also influenced by user behavior, such as the manner in which a user presents a finger, looks at a camera, etc.

**VERIFICATION AND IDENTIFICATION**

Biometric systems can be operated in two modes, named identification and verification. We will refer to recognition for the general case, when we do not want to differentiate between them. However, some authors consider recognition and identification synonymous.

**identification:** In this approach no identity is claimed from the user. The automatic system must determine who the user is. If he/ she belongs to a predefined set of known users, it is referred to as closed-set identification. However, for sure the set of users known (learnt) by the system is much smaller than the potential number of people that can attempt to enter. The more general situation where the system has to manage with users that perhaps are not modeled inside the database is referred to as open-set identification. Adding a "none-of-the-above" option to closed-set identification gives open-set identification. The system performance can be evaluated using an identification rate.

**verification:** In this approach the goal of the system is to determine whether the person is the one that claims to be. This implies that the user must provide an identity and the system just accepts or rejects the users according to a successful or unsuccessful verification. Sometimes this operation mode is named authentication or detection. The system performance can be evaluated using the False Acceptance Rate (FAR, those situations where an impostor is accepted) and the False Rejection Rate (FRR, those situations where a user is incorrectly rejected), also known in detection theory as False Alarm and Miss, respectively. There is a trade-off between both errors, which has to be usually established by adjusting a decision threshold. The performance can be plotted in a ROC (Receiver Operator Characteristic) or in a DET (Detection error trade-off) plot [3]. DET curve gives uniform treatment to both types of error, and uses a logarithmic scale for both axes, which spreads out the plot and better distinguishes different well performing systems and usually produces plots that are close to linear. Note also that the ROC curve has symmetry with respect to the DET, i.e. plots the hit rate instead of the miss probability. DET plot uses a logarithmic scale that expands the extreme parts of the curve, which are the parts that give the most information about the system performance. Figure 1 shows an example of DET of plot, and figure 2 shows a classical ROC plot.

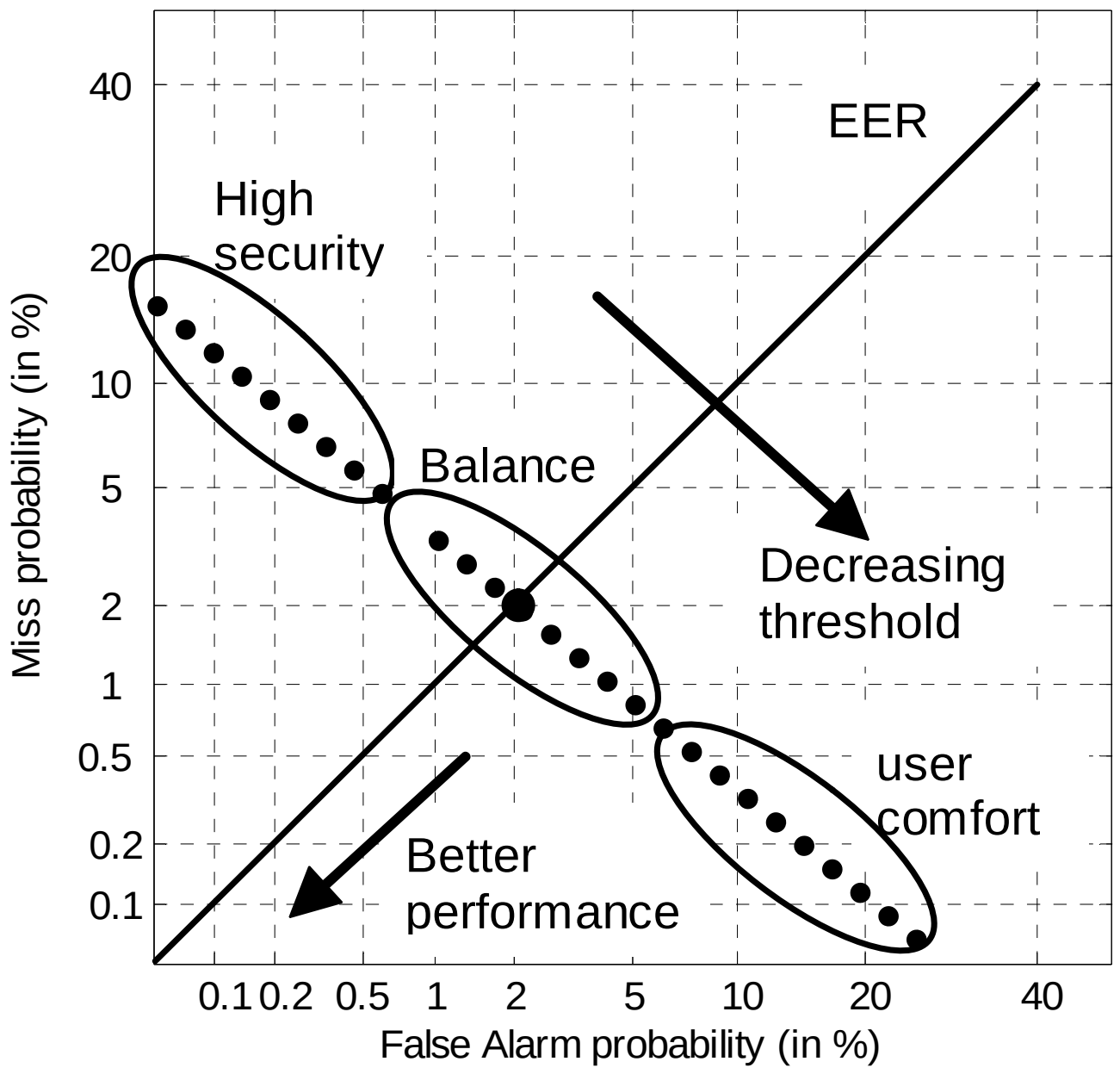

**Figure 1**. Example of a DET plot for a user verification system (dotted line). The Equal Error Rate (EER) line shows the situation where False Alarm equals Miss Probability (balanced performance). Of course one of both errors rates can be more important (high security application versus those where we do not want to annoy the user with a high rejection/ miss rate). If the system curve is moved towards the origin, smaller error rates are achieved (better performance). If the decision threshold is reduced, higher False Acceptance/Alarm rates are achieved.

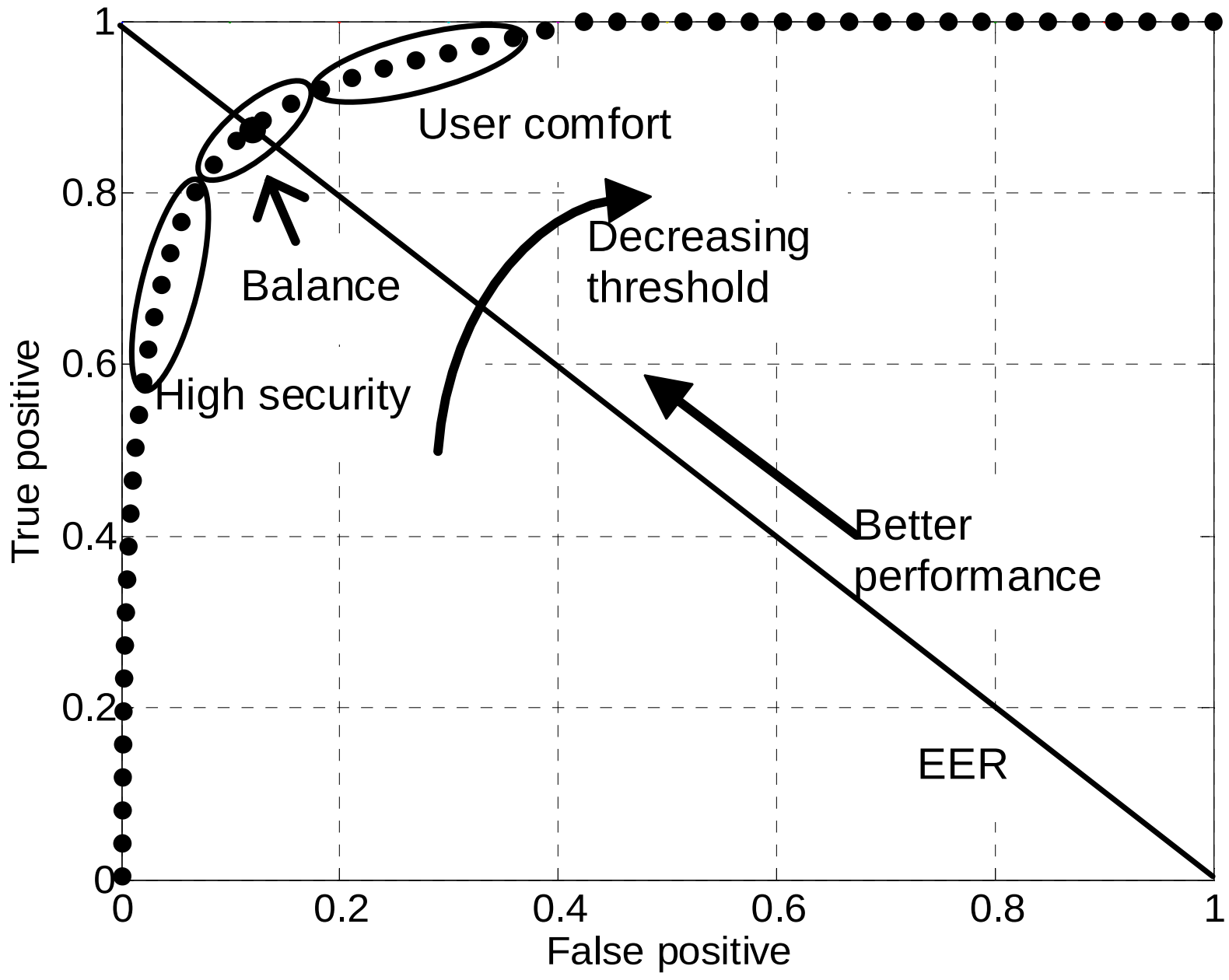

**Figure 2**. Example of a ROC plot for a user verification system (dotted line). The Equal Error Rate (EER) line shows the situation where False Alarm equals Miss Probability (balanced performance). Of course one of both errors rates can be more important (high security application versus those where we do not want to annoy the user with a high rejection/ miss rate). If the system curve is moved towards the upper left zone, smaller error rates are achieved (better performance). If the decision threshold is reduced, higher False Acceptance/Alarm rates are achieved. It is interesting to observe that comparing figures 1 and 2 we get True positive = (1 – Miss probability) and False positive = False Alarm.

For systems working in verification mode, the evolution of FAR and FRR versus the threshold setting is an interesting plot. Figure 6 shows an example of this kind of plot. We can observe that these rates imply opposite requirements. Using a high threshold, no impostor can fool the system, but a lot of genuine users will be rejected. Contrarily, using a low threshold, there would not be inconveniences for the genuine users, but it will be reasonably easy for a hacker to crack the system. According to security requirements, one of both taxes will be more important than the other one.

**Decision error rates versus matching errors**

False Acceptance Rate (FAR) and False Rejection Rate (FRR) include matching error and biometric signal acquisition error. Although they are convenient measures for a potential system user, they have some ambiguity [4] because they vary if the system allows multiple attempts or has multiple templates. For this reason, matching algorithm errors are defined as those for a single comparison of a submitted sample against enrolled template (model). Thus, the following matching errors are defined [4]:

> **False Match Rate (FMR):** It is the expected probability that a sample will be incorrectly declared to match a single randomly-selected "non-self" template.
>
> **False Non-Match Rate (FNMR):** It is the expected probability that a sample will be incorrectly declared not to match a template of the same measure from the same user supplying the sample.

Therefore, false match/ non-match rates are calculated over the number of comparisons; on the other hand, false accept/ reject rates are calculated over transactions and refer to the acceptance or rejection of the stated hypothesis, whether positive or negative. Further, false accept/ reject rates include Failure-To-Acquire rates (defined as the expected proportion of transactions for which the system is unable to capture a biometric signal of adequate quality. The equivalence, when verification decision is based on a single attempt, is the following:

$$FAR = (1–FTA) \times FMR$$
$$FRR = (1–FTA) \times FNMR + FTA$$

**Is identification mode more appropriate than verification mode?**

Certain applications lend themselves to verification, such as PC and network security, where, for instance, you replace your password by your fingerprint, but you still use your login. However, in forensic applications it is mandatory to use identification, because, for instance, latent prints lifted from crime scenes never provide their "claimed identity".

In some cases, such as room access [5-6], it can be more convenient for the user to operate on identification mode. However, verification systems are faster because they just require one-to-one comparison (identification requires one to N, where N is the number of users in the database). In addition, verification systems also provide higher accuracies. For instance, a hacker has almost N times [7] more chance to fool an identification system than a verification one, because in identification he/she just needs to match one of the N genuine users. For this reason, commercial applications operating on identification mode are restricted to small-scale (at most, a few hundred users). Forensic systems [8-9] operate in a different mode, because they provide a list of candidates, and a human supervisor checks the automatic result provided by the machine. This is related to the following classification, which is also associated to the application.

## POSITIVE AND NEGATIVE RECOGNITION

Two kinds of applications can be established, according to the user's attitude [10]:

> **Negative recognition:** the system must establish if the person is the one that (implicitly or explicitly) denies being. The purpose of negative recognition is to prevent a single person from using multiple identities. Classical recognition methods (handheld tokens and knowledge-based) cannot provide this kind of recognition. Thus, only biometrics can be used for negative recognition. Government (ID cards, driving licenses, etc.) and forensic (criminal investigation, terrorist identification, etc.) applications belong to this group, where obviously the user does not want to be identified and will not be collaborative.
>
> **Positive recognition:** those applications that do not require the user to provide his identity, perhaps for convenience (identification requires less operations from the user than verification, so it is easier to operate), belong to this group. Identity verification is also typically used for positive recognition. It is interesting to observe that in this operation mode, the user is interested in being recognized, and will be collaborative. Otherwise, his/ her attempt to enter will be rejected. This clearly contrasts with the previous mode.

Roger Clark summarizes [1] a real case about "false identities". It was reported in 1993, the story of a pensioner who had defrauded the Department of Social Security of $400,000. The defendant had claimed he was born overseas, obviating the need for a birth certificate. He adopted multiple names. For each name, he entered himself on the Electoral Roll, obtained a Tax File Number and submitted a tax return (for amounts which did not incur the payment of a tax), nominated on various application forms the names of various companies as previous employers, visited doctors and acquired various certificates from them, and using these obtained membership of a roadside

assistance association, insurance polices, union membership, Medicare and government concession cards. Obviously, this case can only be detected by means of crossed searches using biometric data.

**BIOMETRIC TECHNOLOGIES**

Several biometric traits have been proven useful for biometric recognition. Nevertheless, the general scheme of a biometric recognition system is similar, in all the cases, to that shown in figure 3.

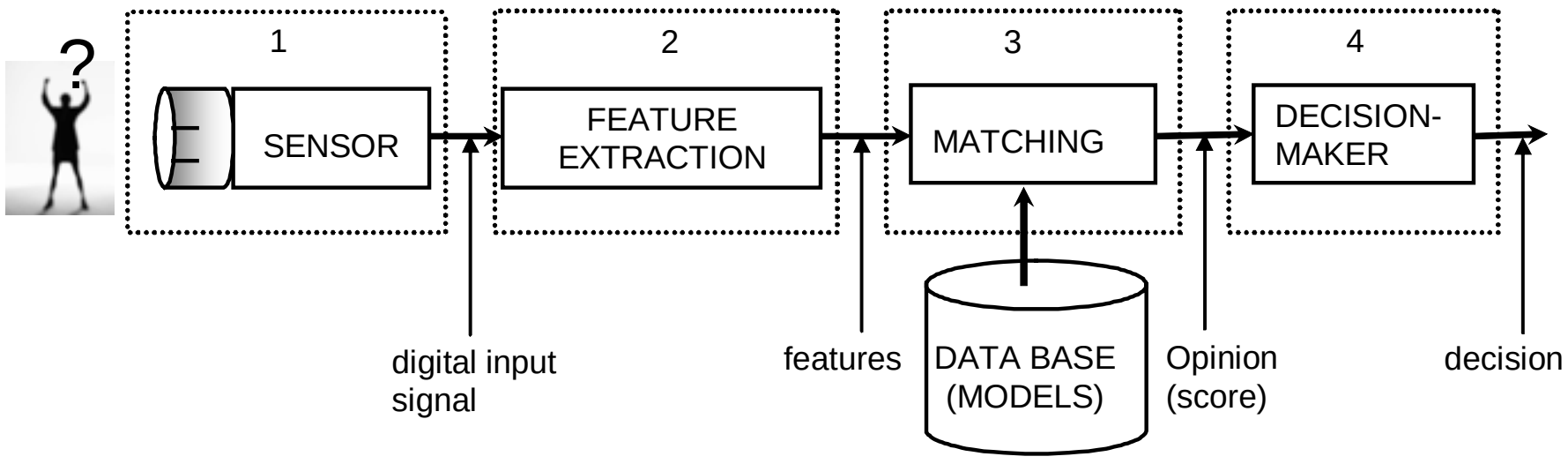

**Fig. 3 General scheme of a biometric recognition system.**

The scheme shown in figure 3 is also interesting for vulnerability study [11] and improvements by means of data fusion [12] analysis. In this paper, we will restrict to block number one, the other ones being related to signal processing and pattern recognition. Although common sense points out that good acquisition is enough for performing good recognition, at least for humans, this is not true. It must be taken into account that next blocks, numbered 2 to 4 in figure 3, are indeed fundamental. A good image or audio recording is not enough. Even for human beings, a rare disorder named agnosia exists. Those individuals suffering agnosia are unable to recognize and identify objects or persons despite having knowledge of the characteristics of the objects or persons. People with agnosia may have difficulty recognizing the geometric features of an object or face or may be able to perceive the geometric features but do not know what the object is used for or whether a face is familiar or not. Agnosia can be limited to one sensory modality such as vision or hearing. A particular case is named face blindness or prosopagnosia [13] (prosopon is a Greek word for face). Prosopagnosics often have difficulty recognizing family members, close friends, and even themselves. They often use alternative routes to recognition, but these routes are not as effective as recognition via the face. Agnosia can result from strokes, dementia, or other neurological disorders. More information about agnosia can be obtained from the National Organization for Rare Disorders (NORD) [14]. Thus, biometric recognition, even when performed by humans, is not a trivial task. It is an extremely complicated process performed by our brain!

Table 2 summarizes some possibilities for different biometric traits acquisition. Obviously, properly speaking, some sensors require a digitizer connected at its output, which is beyond the scope of this paper. We will consider that block number one produces a digital signal which can be processed by a Digital Signal Processor (DSP) or Personal Computer (PC).

| **Biometric trait** | **Sensor** | **Comments** |
|---|---|---|
| Fingerprint | Ink+ paper + scanner | Classical method is becoming old-fashion, because the ink is annoying. However, it can acquire from nail to nail borders, and the other methods, especially when using low-cost devices, provide a limited portion of the fingerprint. |
| | Optical | It is the most widely used and easy-to-operate technology. It can acquire larger surfaces than the capacitive ones, but due to optical phenomenon, it shows some distortions. |
| | Capacitive | They are easy to integrate into small, low-power and, low-cost devices. However, they are more difficult to operate than the optical ones (wet and/ or warm fingers). |
| | Ultrasound | They are not ready for mass-market applications yet. However, they are more capable of penetrating dirt than the other ones, and are not subject to some of the image-dissolution problems found in larger optical devices. |
| | Photo-camera | Nowadays almost all the mobile phones have a photo-camera, enabling face recognition applications. They provide high resolutions and quality. |
| Face | Video-camera | A sequence of images alleviates some problems, such as face detection and offers more possibilities. Although they were expensive some decades ago, currently it is possible to acquire cheap webcams. They produce smaller resolutions than photo-cameras. Thus, theoretically, the user should be closer to the camera than using a photo-camera, in order to obtain the same resolution. |
| Speech | Microphone | The telephone system provides a ubiquitous network of sensors for acquiring speech signals. Even for non-telephone applications, sound cards and microphones are low-cost, easy-to-operate, and readily available. |
| Iris | Kiosk-based systems | The camera searches for eye position. They are the most expensive ones and the easier to operate. |
| | Physical access devices | The device requires some user effort: a camera is mounted behind a mirror. The user must locate the image of his eye within a 1-inch by 1-inch square surface on the mirror. |
| | Desktop cameras | The user must look into a hole and look at a ring illuminated inside. These equipments are the cheapest ones, but they have proven fairly difficult for some users, who are unable to be enrolled on the system. |
| Retina | Retina-scanner | A relative large and specialized device is required. It must be specifically designed for retina imaging. Image acquisition is not a trivial matter. |
| Signature | Ball pen + paper + scanner/ camera | The system recognizes the signature analyzing its shape. This kind of recognition is known as “off-line”, while the other ones are “on-line”. |
| | Graphics tablet | It acquires the signature in real time. Some devices can acquire: position in x and y-axis, pressure applied by the pen, azimuth and altitude angles of the pen with respect to the tablet (see Fig. 11). It is very difficult to replicate the dynamic information (pressure, azimuth, altitude, etc.) for each digitized signature point. Thus, it is safer than the previous one. |
| | PDA | Stylus operated PDAs are also possible. They are becoming more popular, so there are some potential applications. |
| Hand-geometry | Hand-scanning device | Commercial devices consist of a covered metal surface, with some pegs for ensuring the correct hand position. A series of cameras acquire three dimensional (3D) images of the back and the sides of the hand. Obviously they are not general purpose devices (you cannot use them for another matter), and the price is high. |
| | Conventional scanner | Some research groups at universities have developed systems based on images acquired by a conventional document scanner. Thus, the cost is reduced, but the acquisition time is at least 15 seconds per image. They just provide the hand profile, but it is possible to take advantage of the palmprint |
| | Conventional camera | Some research groups at universities have developed systems based on images acquired by conventional cameras. Using some “bricolage” it is possible to obtain a 3D image. They are faster than the previous acquisition method, and do not suffer the problem of palm-print flattening. |
| Palm-print | Document scanner | Although there are not commercial applications, some research groups at universities have developed systems based on images acquired by a conventional document scanner. Resolution must be higher than for hand-geometry, which implies larger acquisition times. |
| Keystroke | keyboard | Although not used habitually, standard keyboards can measure how long keys are held down and duration between key instances, which is enough for recognition. |

Table 2. Some biometric traits and possible acquisition sensors.

Figure 4 shows some biometric traits, and figure 5 some biometric scanners.

<table>
<tr>
<td>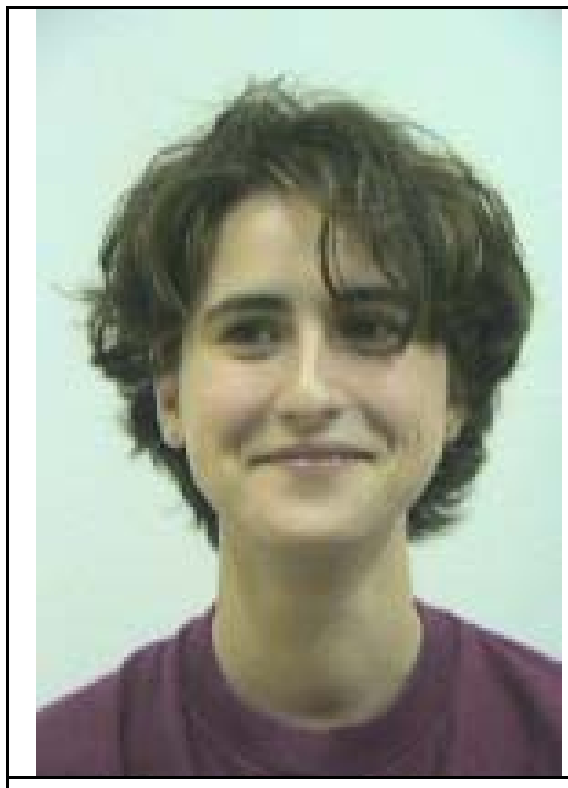</td>
<td>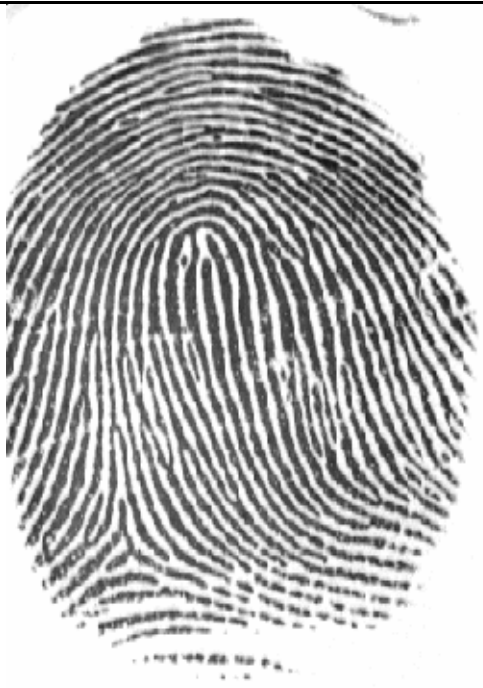</td>
<td>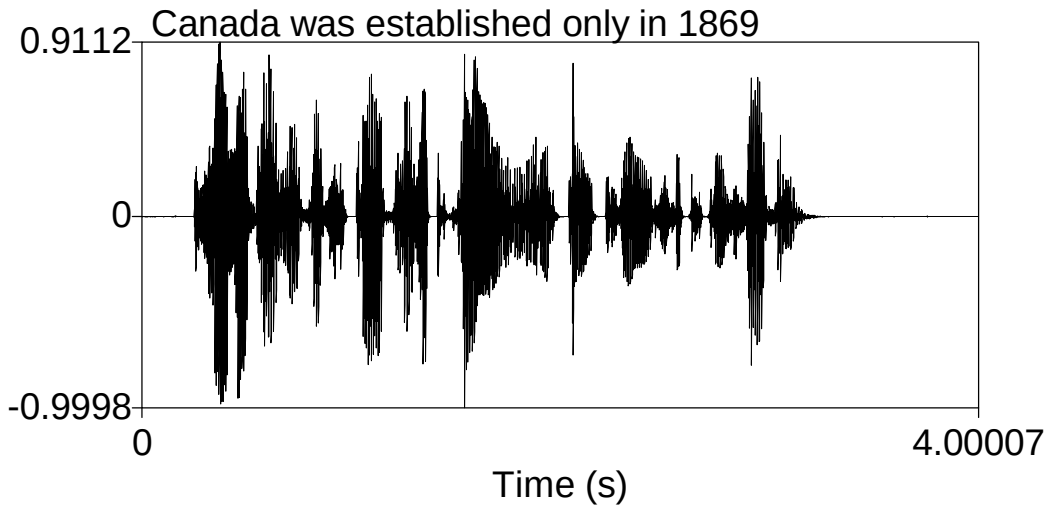
</td>
</tr>
<tr>
<td>Face acquired with digital camera</td>
<td>Fingerprint Acquired with optical sensor</td>
<td>Speech<br>Acquired with a microphone</td>
</tr>
<tr>
<td>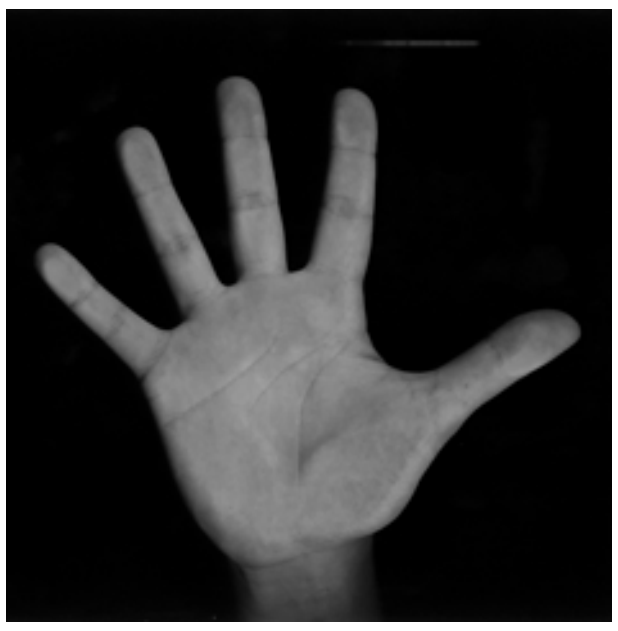</td>
<td>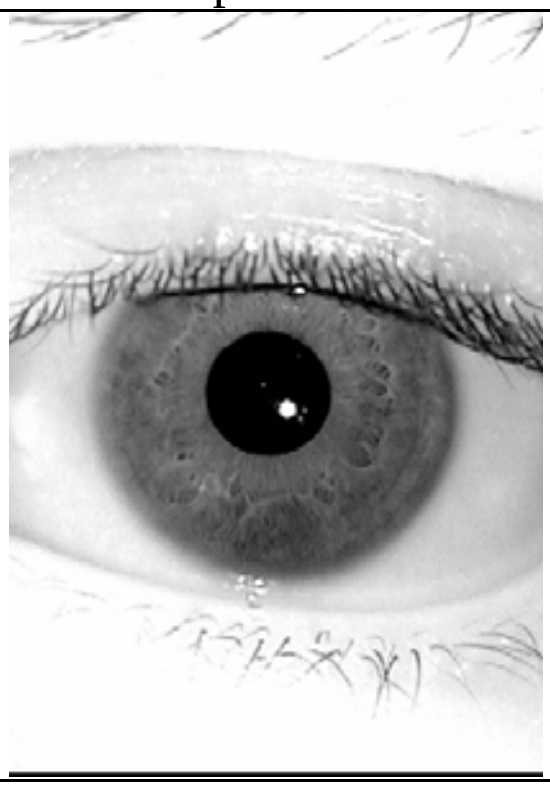</td>
<td>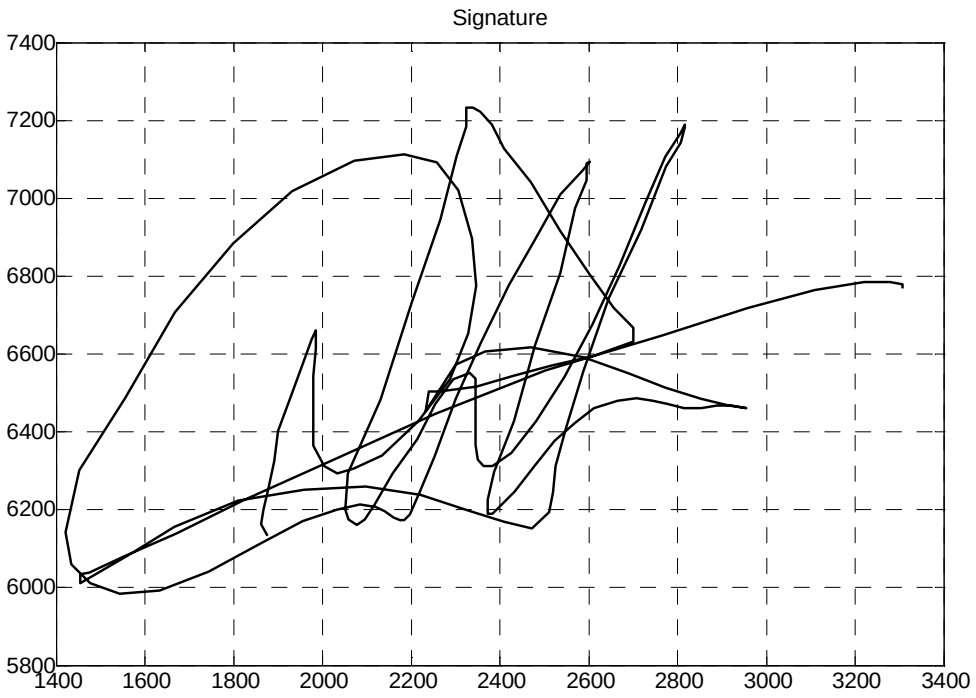
</td>
</tr>
<tr>
<td>2D Hand geometry acquired with a document scanner</td>
<td>Iris acquired with a physical access device</td>
<td>Signature<br>Acquired with a graphics tablet</td>
</tr>
</table>

**Fig. 4. Examples of biometric traits**

<table>
<tr>
<td>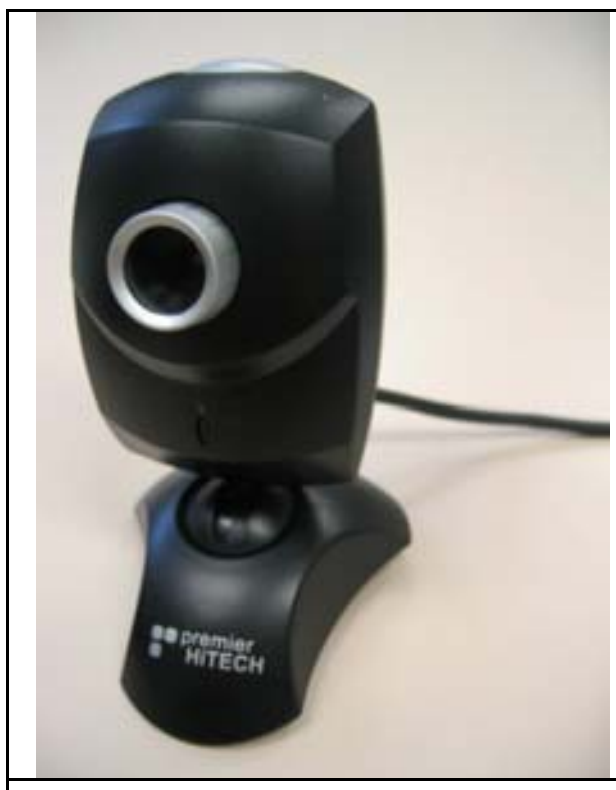
</td>
<td>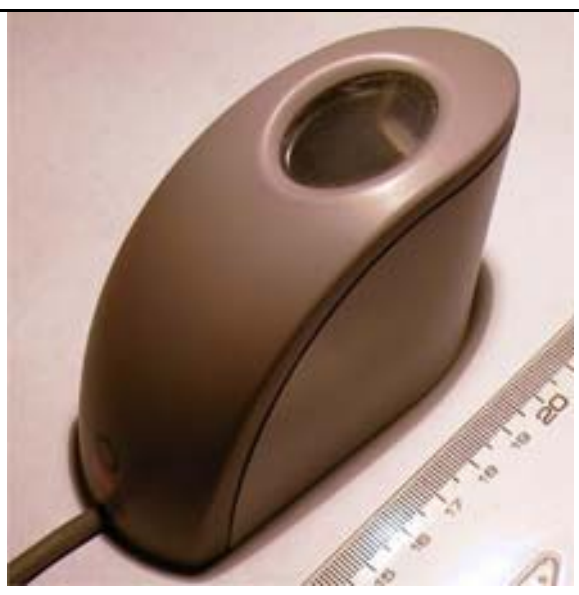</td>
<td>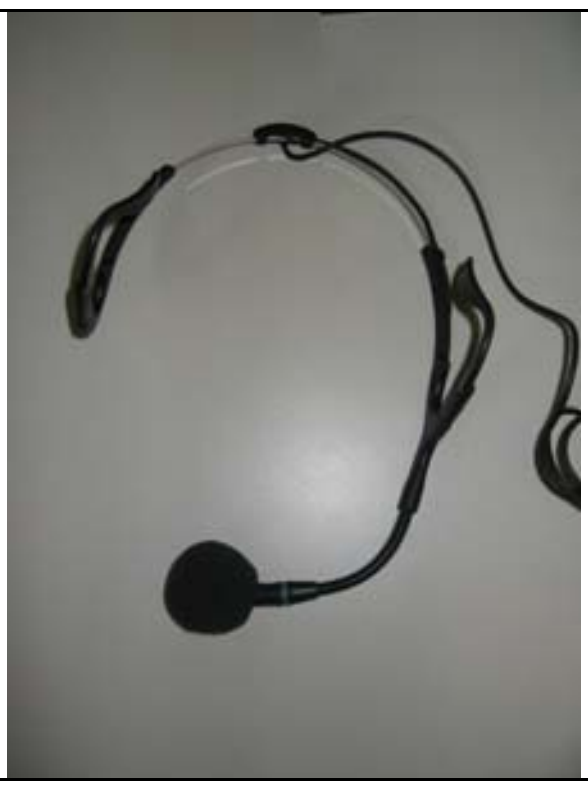</td>
</tr>
<tr>
<td>Webcam</td>
<td>Fingerprint optical scanner</td>
<td>Headset microphone for speech</td>
</tr>
<tr>
<td>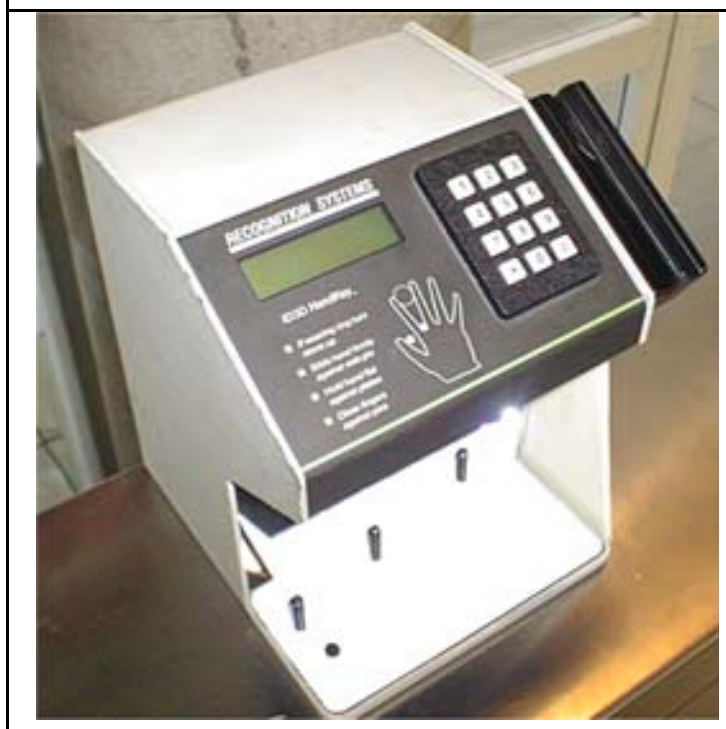</td>
<td>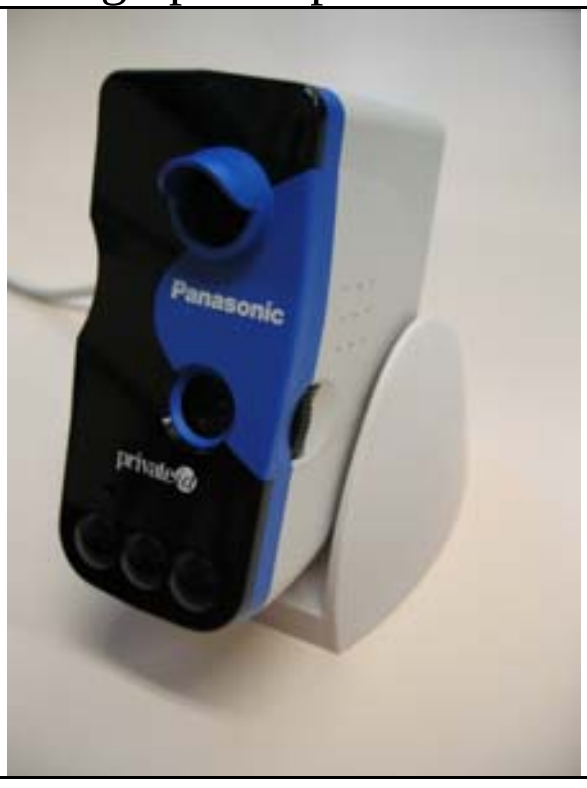
</td>
<td>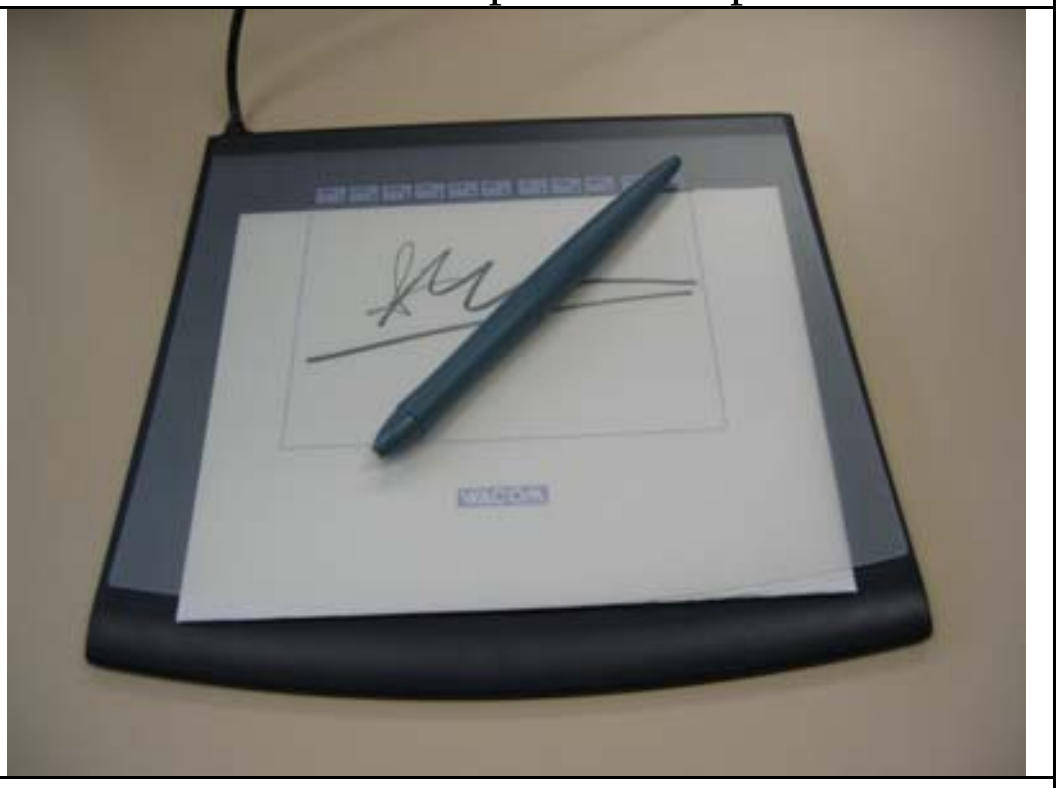</td>
</tr>
<tr>
<td>3D Hand-geometry scanner</td>
<td>Iris Desktop camera</td>
<td>Graphics tablet for signature</td>
</tr>
</table>

**Fig. 5. Some commercial biometric scanners.**

**How does a biometric system work?**

Usually biometric systems require two consecutive steps:

**Enrollment:** In a similar fashion as humans do, the system needs a learning procedure, before being able to recognize (it is obviously hard to recognize a person that has not been "seen" before). The purpose of enrollment is to have user's characteristics registered for later use. The procedure consists of the following steps:

a) The input signal is acquired (block 1 in figure 3) by means of a biometric scanner. If possible, the quality of the sensed signal is checked. If it is below a threshold, a new acquisition is performed.
b) Some measurements are extracted from this signal (block 2 in figure 3) by means of digital signal processing.
c) Measured parameters of previous step are used to work out a model for the given user. Some times, the whole set of extracted features are stored, and used as model. This model will be compared with the features extracted from input signals on recognition mode.

The proportion of individuals for whom the system is unable to generate repeatable templates is defined as Failure to Enroll (FTE) rate. FTE includes those unable to present the required biometric feature (for example an Iris system can fail to enroll the iris of a blind eye), those unable to produce an image of sufficient quality at enrolment, as well as those unable to reproduce their biometric feature consistently (the system cannot reliably match their template in attempts to confirm that the enrollment is usable).

**Recognition:** Once the user is enrolled, the system can work in identification or verification mode, using the block diagram shown in figure 3, and the systematic explained in previous sections (one-to-many comparisons for identification, and one-to-one for verification using the claimed identity by the user).

Each technology presents differences. For this reason, a short summary of the most successful ones is included.

**Face**

Face recognition is probably the most natural way to perform a biometric authentication between human beings. Face recognition can rely on single still images, multiple still images, or video sequence. Although traditionally most efforts have been devoted to the former one, the latest ones are quickly emerging [15], probably due to the reduction of price in image and video acquisition devices. For instance, a sequence of images can provide a unimodal data fusion scheme, where the verification relies on a set of images, rather than on a single one. Figure 6 shows an example [16] where each test consists of a single still image (on the left) and the best match of five still images. We can observe an improvement on the FRR with a minor degradation on FAR (FRR plot shifts to the right in larger amount than FAR, yielding more separation between plots and less critical tradeoff for threshold setting up). This is similar to the PIN keystroke on ATM cashiers, where three attempts are offered. This strategy avoids inconveniences for users, with a negligible degradation on PIN vulnerability. Obviously, it can also be applied to other biometric traits. However, a video-camera lets to easily obtain a consecutive sequence of images in a short time period. For fingerprints, for instance, it would not have too much sense, and it would be time consuming, to ask the user for five consecutive acquisitions.

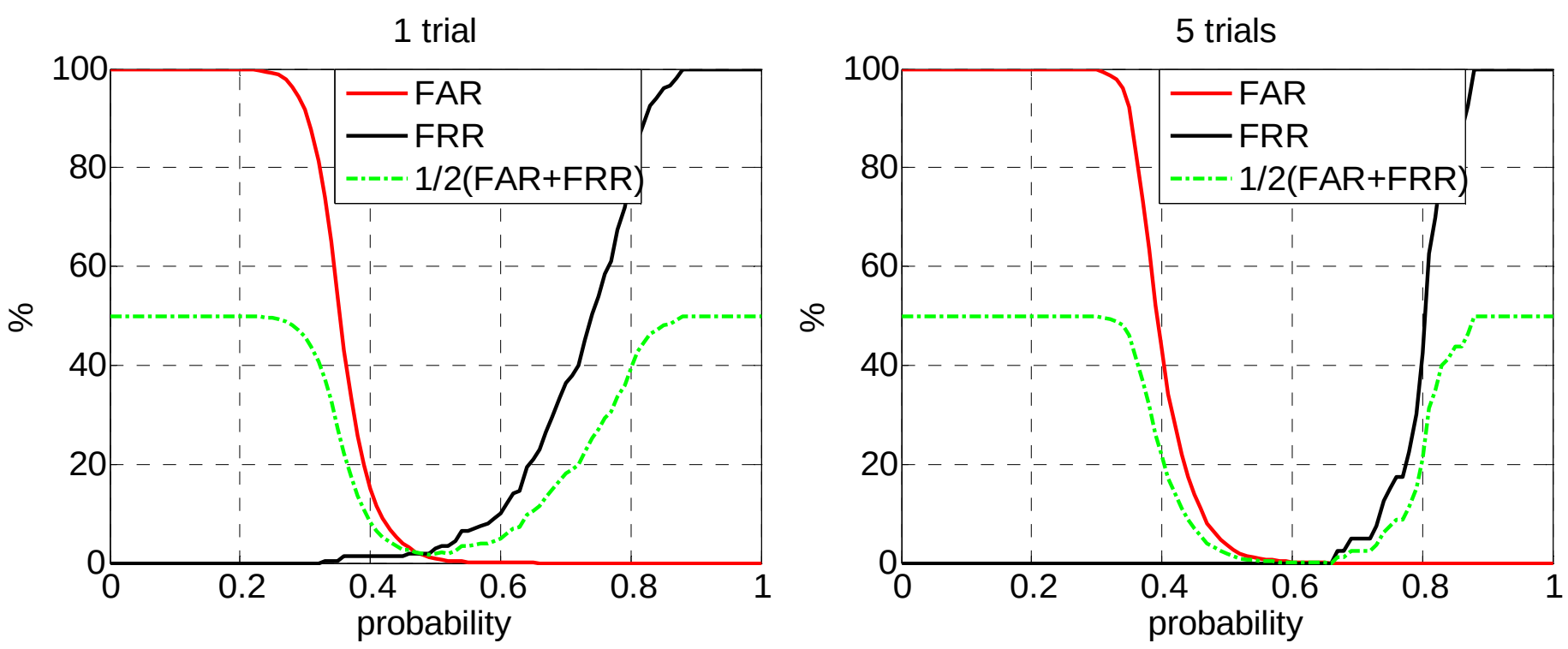

**Fig. 6.** FAR and FRR offering 1 and 5 trials.

**Fingerprint**

Fingerprint recognition looks at the patterns found on a fingertip. Several algorithms exist. Interested reader can find more details in [7]. The use of fingerprints as a biometric is both the oldest mode of computer-aided recognition, and the most widely established today. It has been estimated that the probability to find two persons with the same fingerprint is one in one billion.

However, some users can be reluctant to use it. In relatively free countries, the situation generally is that [1] there is no authority for the compulsory provision of fingerprints, unless they are being charged with a criminal offence; and there is no authority for the prints to be retained unless the charge is pursued and the offence proven. Thus, fingerprints have some common-criminal stigma. However, recently, fingerprinting is being applied for immigration matters, and gaining acceptance as recognition method not necessarily related to criminals.

**Speech**

Speech signals can be easily acquired, as stated on table 2. However, one of the critical facts for speaker recognition is the presence of channel variability from training to testing, that is, different signal-to-noise ratio, kind of microphone, evolution with time, etc. For human beings this is not a serious problem, because of the use of different levels of information. However, this affects automatic systems in a significant manner. Fortunately higher-level cues are not as affected by noise or channel mismatch. Some examples of high-level information in speech signals are speaking and pause rate, pitch and timing patterns, idiosyncratic word/phrase usage, idiosyncratic pronunciations, etc. Current systems try to take advantage of these different levels [17]. In addition, when compared to other biometric traits, speech offers more possibility, because the system can ask the user for a specific input sentence. This is known as text-dependent mode. In this way, a previous recording of a genuine speech will not be accepted, because it will not match the desired sentence, which can belong to a relative large set of possible sentences. This is in contrast with fingerprints, for instance, where the system just can ask the user to put his/her finger on the surface, or perhaps to ask for a specific finger, which will just be one of ten possibilities.

**Hand-geometry**

Hand-geometry measures the shape of the hand. This technology provides small template size (from 9 to 25 bytes) and low storage needs. Feature extraction is relatively simply, so low-cost processors can be used. It can be as simple as contour extraction with the scheme [18] shown in figure 7, and measurements such as lengths, perimeters and areas. Figure 8 shows some measured features.

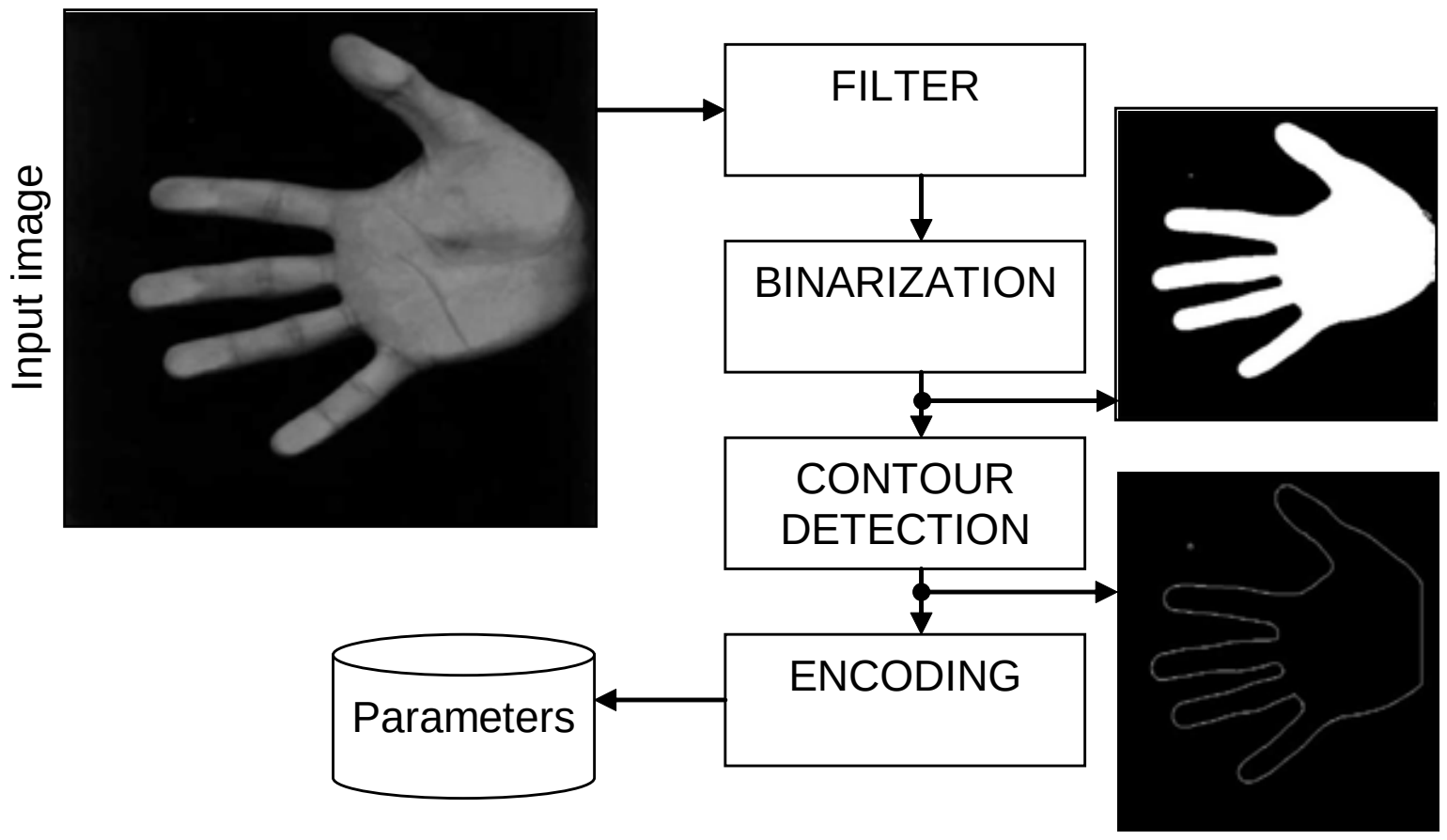

**Fig.** 7. Block diagram for 2D hand-geometry recognition.

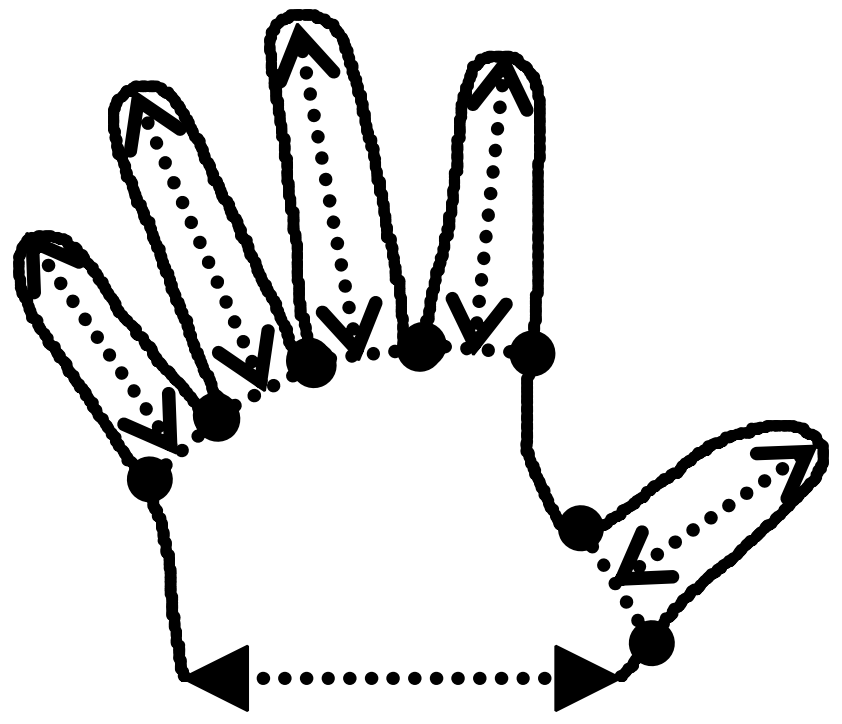

**Fig. 8** measured features for 2D hand-geometry

In addition, it is possible to perform a palm-print recognition, which takes advantage of the inner surface between the wrist and the fingers. This region presents a rich information, including wrinkles and ridges/valleys similar to a fingerprint, but in a larger surface. Although there are not commercial systems using palm-prints, some experiments have been performed at universities using this information alone, and combined with hand-geometry for improved accuracy.

**Iris**

Iris recognition offers a very high capability to distinguish between individuals, even between user's left and right eyes. It has been estimated that the probability that two irises would be identical by random chance is approximately $10^{-35}$ [19]. It is even higher than fingerprints. However, it is difficult to use. Especially when using the lowest-cost devices. Figure 9 shows the user interaction with the system for a desktop camera, with the help of a measure. The distance between user and camera must be around half a meter. Then, the user must center his eye in order to see a ring inside the camera aperture. Some systems provide a spot of light: when the light fills up the circle, it means that the user is at the correct distance. If the light is larger than the circle, he/ she is too close. If the light does not fill the circle completely, this means that he/ she is too far. Other cameras provide a luminous circle that turns, for instance, from orange to green when the user is properly set. These last systems are more complicate to operate, because you just receive binary information: correct/ incorrect, but you do not know the correct movement to perform. In addition, you should show a naked eye, so in most of the cases (depending on shadows and reflexes) you must take your glasses off, and perhaps to use a measure that help you to achieve the required distance (see figure 9).

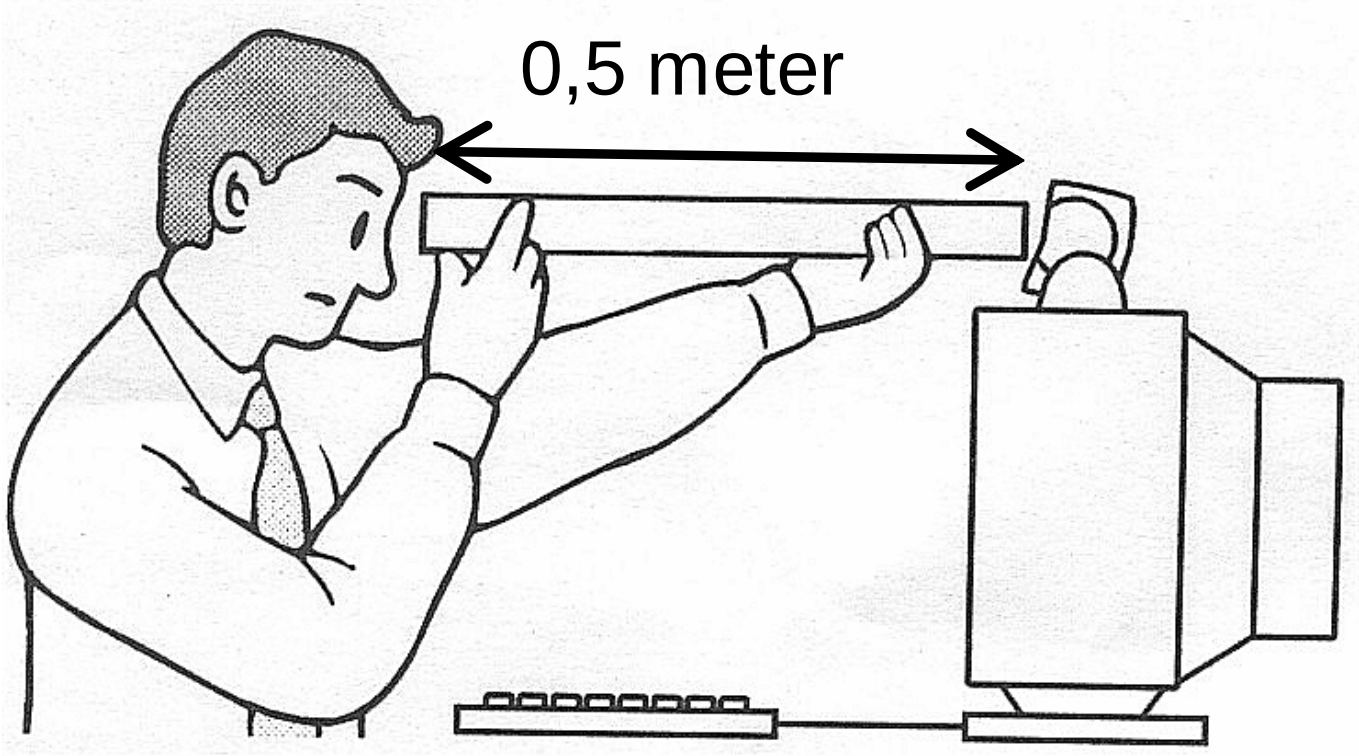

**Fig. 9** User interaction with a desktop iris camera.

Some users experiment problems with Iris scanners, such as eastern people, because the eyelid can occlude the iris. In this case, some manual actuation must be done in order to remove the occlusion (see figure 10).

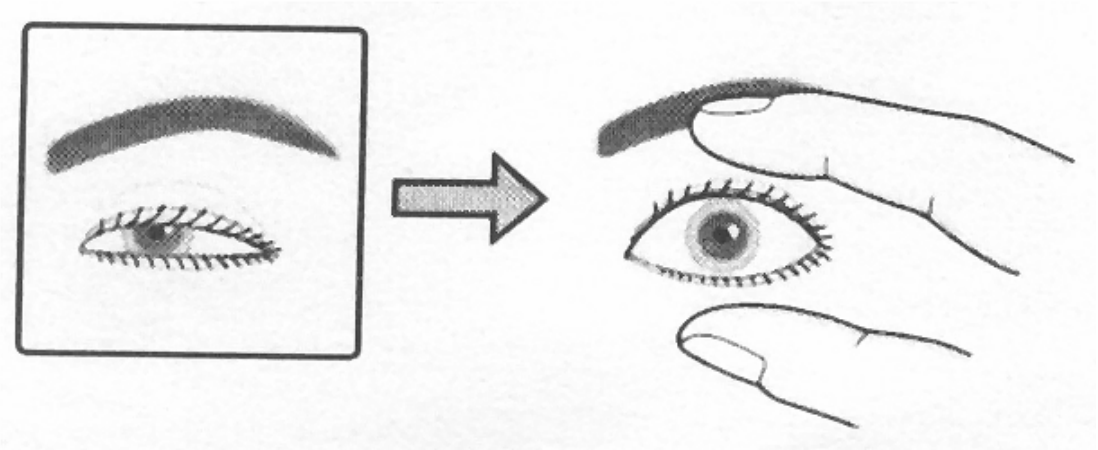

**Fig. 10** Illustration of possible failure to acquire problems for some users.

**Signature**

Signature is probably the most accepted method for recognition. We frequently use it when signing credit card receipts, checks, etc. In addition, biometrics presents a serious drawback when compared with classical methods passwords and tokens (while it is possible to obtain a new card number, it is not possible to replace any biometric data, which should last forever) [11]. However, signature is an exception, because users can change their signature. Although, theoretically, a person could learn to sign in exactly the same manner as another person, in practice, it is very difficult to replicate the dynamic information (pressure, azimuth, altitude, etc., see fig. 11) for each digitized signature point (pixel), which cannot be ascertained from examining a written signature or by observing a person signing. For a more complete description of signature recognition, you can read [20].

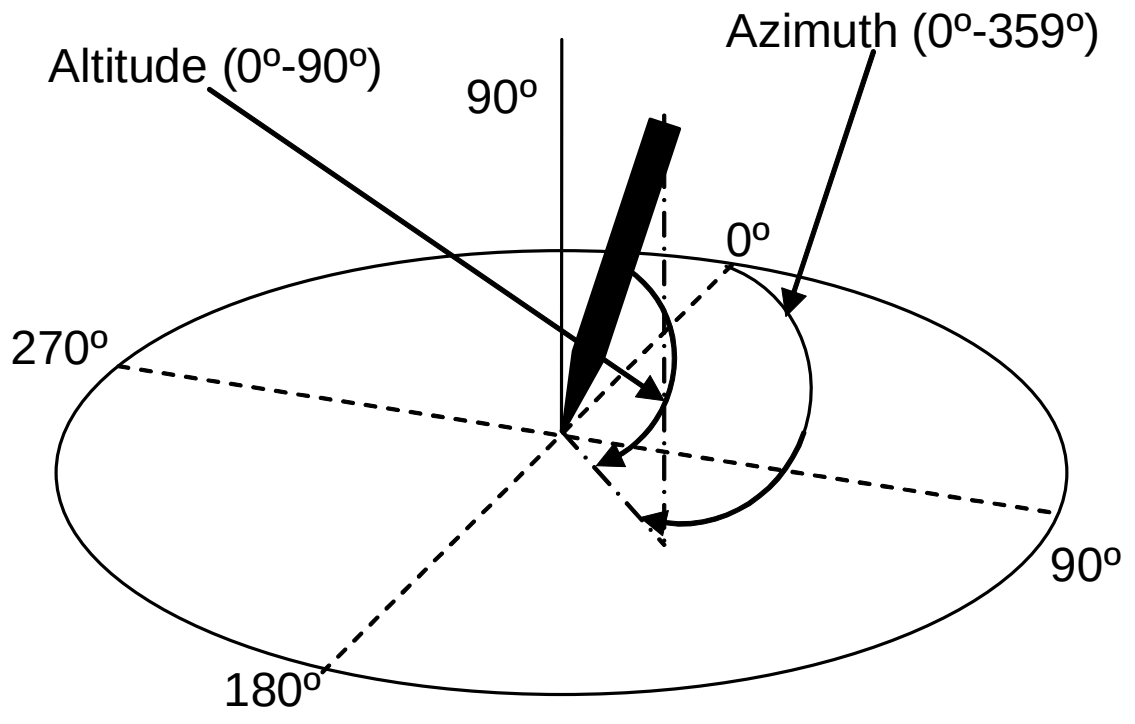

**Fig.11 Azimuth and inclination angles of the pen with respect to the plane of the graphic card**

**Retina**

This technology involves analyzing the layer of blood vessels situated at the back of the eye. For this purpose, a low-intensity light source must enter through the pupil. It is extremely accurate, but requires the user to look into a receptacle and focus on a given point. Thus, although it works well, it has not been accepted by users, and applications are restricted to high-level government, military, and corrections applications.

Table 3 compares some properties of different biometric technologies. It is interesting to observe that nobody can claim that a given technology outperforms the other ones for all the aspects. Table 4 summarizes the main weaknesses and critical situations for each technology.

| Characteristic | face | fingerprint | speech | hand | iris | Signature |
|---|---|---|---|---|---|---|
| Ease of use | M/H | H | H | H | L/M | H |
| Accuracy | M | H | M | M | H | M/L |
| Acceptability | H | M | H | M/H | L/M | H |
| Security | M | H | M | M | H | M/L |
| Permanence | M | H | M/L | M/H | H | M/L |

**Table 3. Comparison of different biometric technologies (L=Low, M=Medium, H=High)**

| Biometric technology | Weaknesses |
|---|---|
| Fingerprint | ▪ Certain users do not have fingerprints (elder people, some Asian populations, manual workers with acid, cement, etc.)<br>▪ Some fingerprint scanners cannot acquire fingerprints that are too oily, dry, wet, warm, etc.<br>▪ Temporal or permanent damages can make fingerprint recognition impossible. |
| Face | ▪ Changes in hairstyle, makeup, facial hair, etc.<br>▪ Addition or removal of glasses, hats, scarf, etc.<br>▪ Dramatic variations of weight, skin color change due to sun exposure, etc. |
| Iris | ▪ Eye trauma is rarely present, but still possible. Although this system is quite robust, it is not popular nor the sensors are widely-introduced. |
| Voice | ▪ Illness can modify the voice (cold, flu, aphonia, etc.)<br>▪ Acquisition devices and environments can vary significantly, for instance in mobile phone access. This degrades the recognition rates. |
| Hand geometry | ▪ Weight increases or decreases, injuries, swelling, water retention, etc. can make recognition impossible.<br>▪ Some users can be unable to locate the hand correctly due to paralysis, arthrosis, etc. |

**Table 4. Drawbacks of the main biometric systems.**

Another important issue is the accuracy of each technology. For this purpose, it is more reliable to look at the figures of well established competitions, where a group of universities and private companies evaluate their own technology using a common database. This makes feasible the direct comparison between competing algorithms and lets to provide a realistic number about the state-of-the-art for those evaluated technologies. However, we must take into account that, in general, each experiment has a different set of users, so it is not possible to compare the verification errors of a same population, using different technologies. It is just an approach of each technology.

Table 5 summarizes the results of some studies, such as FVRT (Face Recognition Vendor Test) [21], CESG (Communications Electronics Security Group) [22], FVC (Fingerprint Verification Competition) [23] and NIST (National Institute of standard technologies) [24], and SVC (Signature Verification Competition) [25]. This table is an update of [26]. A nice property of CESG evaluation is that all the results have been obtained with the same set of 200 users.

| biometric | Test | Test parameter | Attempts | FRR | FAR | FTE | FTA |
|---|---|---|---|---|---|---|---|
| Face | FRVT | 11-13 months spaced | 1 | 4% | 10% | - | - |
| | CESG | 200 users, 1-3 months spaced | 3 | 6% | 6% | 0.0% | 0.0% |
| Fingerprint | FVC | 100 users, Mainly age 20-30 | 1 | 2% | 0.02% | - | - |
| | CESG | 200 users, Mainly age > 25 | 3 | 2% | 0.01% | 1%–2% | 0.4%–2.8% |
| Hand | CESG | 200 users, Mainly age > 25 | 1 | 3% | 0.3% | 0.0% | 0.0% |
| | CESG | 200 users, Mainly age > 25 | 3 | 1% | 0.15% | 0.0% | 0.0% |
| Iris | CESG | 200 users, Mainly age > 25 | 1 | 2% | 0.0001% | 0.5% | 0.0% |
| | CESG | 200 users, Mainly age > 25 | 3 | 0.25% | 0.0001% | 0.5% | 0.0% |
| Voice | NIST | Text independent | 1 | 7% | 7% | - | - |
| | CESG | Text dependet | 3 | 2% | 0.03% | 0.0% | 2.5% |
| Signature | SVC | 60 users, skilled forgeries | 1 | 2.89% | 2.89% | - | - |

**Table 5.** Verification error rate pairs chosen from results of benchmark testing for several biometrics and from different tests. FRR= False Rejection Rate, FAR= False Acceptance Rate, FTE= Failure To Enroll rate, FTA= Failure To Acquire rate.

## SECURITY AND PRIVACY

A nice property of biometric security systems is that security level is almost equal for all users in a system. This is not true for other security technologies. For instance, in an access control based on password, a hacker just needs to break only one password among those of all employees to gain access. In this case, a weak password compromises the overall security of every system that user has access to. Thus, the entire system's security is only as good as the weakest password [10]. This is especially important because good passwords are nonsense combinations of characters and letters, which are difficult to remember (for instance, "Jh2pz6R+"). Unfortunately, some users still use passwords such as "password", "Homer Simpson" or their own name.

Although biometrics offers a good set of advantages, it has not been massively adopted yet [27]. One of its main drawbacks is that biometric data is not secret and cannot be replaced after being compromised by a third party. For those applications with a human supervisor (such as border entrance control), this can be a minor problem, because the operator can check if the presented biometric trait is original or fake. However, for remote applications such as internet, some kind of liveliness detection and anti-replay attack mechanisms should be provided. This is an emerging research topic. As a general rule, concerning security matters, a constant update is necessary in order to keep on being protected. A suitable system for the present time can become obsolete if it is not periodically improved. For this reason, nobody can claim that has a perfect security system, and even less that it will last forever.

Another interesting topic is privacy, which is beyond the scope of this paper. It has been recently discussed in [28].